\documentclass{article}

\usepackage{ICASSP2021}
\usepackage[colorinlistoftodos,prependcaption,textsize=tiny]{todonotes}
\usepackage[nolist,nohyperlinks]{acronym}
\usepackage{cite}
\usepackage{lipsum}
\usepackage{mathtools}
\usepackage{hyperref}
\usepackage{booktabs}
\usepackage{multirow}
\usepackage{array}
\usepackage{amssymb}
\usepackage{algorithm}
\usepackage[noend]{algpseudocode}

\newcolumntype{P}[1]{>{\centering\arraybackslash}p{#1}}
\graphicspath{{./pictures/}}

\title{Regularized Forward-Backward Decoder for Attention Models}
\name{Tobias Watzel, Ludwig Kürzinger, Lujun Li, Gerhard Rigoll}
\address{
  Chair of Human-Machine Communication, Technical University of Munich}

\begin{document}
	
\begin{acronym}[ASR]
	\acro{asr}[ASR]{automatic speech recognition}
	\acro{am}[AM]{acoustic model}
	\acro{lm}[LM]{language model}
	\acro{pm}[PM]{pronunciation model}
	\acro{ctc}[CTC]{connectionist temporal classification}
	\acro{s2s}[Seq2Seq]{sequence-to-sequence}
	\acro{nn}[NN]{neural network}
	\acro{rnn}[RNN]{recurrent neural network}
	\acro{lstm}[LSTM]{long short-term memory}
	\acro{blstmp}[BLSTMP]{bidirectional long short-term memory projected}
	\acro{hmm}[HMM]{hidden markov model}
	\acro{l2r}[L2R]{left-to-right}
	\acro{r2l}[R2L]{right-to-left}
	\acro{bpe}[BPE]{byte pair encoding}
	\acro{dtw}[DTW]{dynamic time warping}
	\acro{cnn}[CNN]{convolutional neural network}
	\acro{wer}[WER]{word error rate}
	\acrodef{char}[char]{character}
\end{acronym}

\maketitle
 
\begin{abstract}
Nowadays, attention models are one of the popular candidates for speech recognition. So far, many studies mainly focus on the encoder structure or the attention module to enhance the performance of these models. However, mostly ignore the decoder. In this paper, we propose a novel regularization technique incorporating a second decoder during the training phase. This decoder is optimized on time-reversed target labels beforehand and supports the standard decoder during training by adding knowledge from future context. Since it is only added during training, we are not changing the basic structure of the network or adding complexity during decoding. We evaluate our approach on the smaller TEDLIUMv2 and the larger LibriSpeech dataset, achieving consistent improvements on both of them.
\end{abstract}

\section{Introduction}
Automatic speech recognition\acused{asr} (\ac{asr}) systems have increased their performance steadily over the years. The introduction of \acp{nn} into the area of speech recognition led to various improvements. Hybrid approaches replaced traditional Gaussian mixture models by learning a function between the input speech features and \acused{hmm}\acl{hmm} states in a discriminative fashion. However, these approaches are composed of several independently optimized modules, i.e., an \acl{am}, a \acl{pm}, and a \acl{lm}. As they are not optimized jointly, useful information cannot be shared between them. Furthermore, specific knowledge is necessary for each module to retrieve the optimal result. 

Recently, \ac{s2s} models are gaining popularity in the community\,\cite{graves2006connectionist, graves2014towards, sutskever2014sequence, tuske2019advancing, weng2018improving, chan2016listen, chiu2018state, chorowski2014end, bahdanau2016end, bahdanau2014neural, graves2012sequence, graves2013speech, sak2017recurrent} since they fuse all aforementioned modules into a single end-to-end model, which directly outputs \acp{char}. Works like \cite{chiu2018state, tuske2019advancing} have already shown that \ac{s2s} models can be superior to hybrid systems\,\cite{chiu2018state} if enough data is available. 
\ac{s2s} models can be categorized into approaches based on \ac{ctc}\,\cite{graves2006connectionist, graves2014towards}, on transducer\,\cite{graves2012sequence, graves2013speech, sak2017recurrent} and on attention\,\cite{sutskever2014sequence, tuske2019advancing, weng2018improving, chan2016listen, chiu2018state, chorowski2014end, bahdanau2016end, bahdanau2014neural}. 

In \ac{ctc}, a \ac{rnn} learns alignments between unlabeled input speech features and a transcript. The basic idea is to assume the conditional independence of the outputs and marginalize over all possible alignments\,\cite{graves2006connectionist}. For \ac{asr}, this assumption is not valid, as consecutive outputs are highly correlated. Transducer models relax the conditional independence and add another \ac{rnn} to learn the dependencies between all previous input speech features and the output\,\cite{graves2013speech}. Attention models also combine two \acp{rnn} with an additional attention network. One \ac{rnn} acts as an encoder to transform the input data into a robust feature space. The attention model creates a glimpse given the last hidden layer of the encoder, the previous time-step attention vector and the previous time-step decoder output. The decoder \ac{rnn} then utilizes the glimpse and the previous decoder output to generate chars\,\cite{bahdanau2014neural}.

In our work, we propose a novel regularization technique by utilizing an additional decoder to improve attention models. This newly added decoder is optimized on time-reversed labels. Since we primarily focus on improving the training process, we utilize the decoder only during the optimization phase and discard it later in the inference. Thus, the network architecture of a basic attention model is not changed during decoding. A recent study demonstrated that it is beneficial to add a \ac{r2l} decoder to a conventional \ac{l2r} decoder\,\cite{mimura2018forward}. The \ac{r2l} decoder is trained on time-reversed target labels and acts as a regularizer during optimization. Their work focused mainly on the advantage of using the additional information to improve the beam search in decoding. They applied a constant scalar value, which attached a more significant weight on the loss function of the standard \ac{l2r} decoder. Furthermore, they trained their models on Japanese words whereby label and time-reversed label sequences were equal. Another comparable work has been published in the domain of speech synthesis. In\,\cite{zheng2019forward}, they also utilized a second \ac{r2l} decoder, combined both losses and added another regularizing function for the \ac{l2r} and \ac{r2l} decoder outputs. Similar to\,\cite{mimura2018forward}, they trained only on equal sequence lengths. In the English language, however, \acp{bpe} for encoding the target transcripts seem superior\,\cite{chiu2018state, tuske2019advancing}. As encoding a time-reversed transcript produces unequal sequence lengths between \ac{l2r} and \ac{r2l} decoders, regularization of these sequences is challenging. To the best of our knowledge, an in-depth study on how to solve this problem and leveraging the newly added decoder during the optimization process has not been done for attention models. Our contributions are to introduce an optimization scheme inspired by \cite{zheng2019forward} for attention models in \ac{asr} and utilize the added decoder during the training. Furthermore, we propose two novel regularization terms for equal and unequal output sequence lengths and demonstrate their superiority over conventional attention models.

%
%
%
%
%
%

\section{Proposed Method}

\subsection{Attention Model}

The standard attentional \ac{s2s} model contains three major components: the encoder, the attention module and the decoder. Let ${\boldsymbol{X}=(\boldsymbol{x}_1, \cdots, \boldsymbol{x}_t , \cdots, \boldsymbol{x}_T)}$ be a given input sequence of $T$ speech features and let ${\boldsymbol{y}=(y_1, \cdots, y_k , \cdots, y_{K})}$ be the target output sequence of length $K$. The encoder transforms the input sequence into a latent space:
\begin{equation}
\begin{split}
\boldsymbol{H}^{\text{enc}} & = (\boldsymbol{h}_1^{\text{enc}}, \cdots, \boldsymbol{h}_t^{\text{enc}}, \cdots, \boldsymbol{h}_T^{\text{enc}}) \\
& =\text{Encoder}(\boldsymbol{x}_1, \cdots, \boldsymbol{x}_t, \cdots, \boldsymbol{x}_T),
\end{split}
\end{equation}
where $\boldsymbol{H}^{\text{enc}}$ encodes essential aspects of the input sequence, i.e., characteristics of the speech signal. The resulting hidden encoder states $\boldsymbol{H}^{\text{enc}}$ and the hidden decoder state $\boldsymbol{h}_{k-1}^{\text{dec}}$ are fed into the attention module to predict proper alignments between the $t\text{-th}$ input and $k\text{-th}$ output sequences:
\begin{equation}
\begin{split}
\alpha_{k,t} & = \text{Attention}(\boldsymbol{h}_{k-1}^{\text{dec}}, \boldsymbol{H}^{\text{enc}}) \\
& = \exp(e_{k,t}) / \sum_{t'=1}^{T} \exp(e_{k,t'}),
\end{split}
\end{equation}
where $\boldsymbol{\alpha}_{k} = (\alpha_{k,1},\cdots,\alpha_{k,t})$ are the attention weights and $e_{k,t}$ is the output of a scoring function:

\begin{equation}
e_{k,t} = \text{Scoring}(\boldsymbol{h}_{k-1}^{\text{dec}}, \boldsymbol{h}_t^{\text{enc}}, \boldsymbol{\alpha}_{k-1}).
\end{equation}

Depending on the task, there are several ways to implement scoring functions. We choose the content-based and location-aware attention from\,\cite{chorowski2015attention} for scoring. Based on the attention weights $\boldsymbol{\alpha}_{k}$, a context vector $\boldsymbol{c}_k$ is created to summarize all information in the hidden states of the encoder for the current prediction:
\begin{equation}
\boldsymbol{c}_k = \sum_{t}\alpha_{k,t}\boldsymbol{h}_t^{\text{enc}}.
\end{equation}
The decoder generates the output distribution using the context vector $\boldsymbol{c}_k$ and the decoder hidden state $\boldsymbol{h}_{k-1}^{\text{dec}}$:
\begin{equation}
p(y_k|p_{1:k-1}, \boldsymbol{X}) \sim \text{Generate}(\boldsymbol{h}_{k-1}^{\text{dec}}, \boldsymbol{c}_k),
\end{equation}
where $\boldsymbol{h}_{k-1}^{\text{dec}}$ is a recurrency, usually a \ac{lstm}:
\begin{equation}
\boldsymbol{h}_{k}^{\text{dec}} = \text{LSTM}(\boldsymbol{h}_{k-1}^{\text{dec}}, \boldsymbol{c}_k, y_{k-1}),
\end{equation}
with $y_{k-1}$ being the predicted target label of the previous prediction step. The resulting model is optimized by cross-entropy loss $\mathcal{L}_{\text{CE}}$.

\subsection{Adding a Backward Decoder}
For a traditional attention model, the char distribution $p(\overrightarrow{y}_k|\newline p_{1:k-1}, \boldsymbol{X})$ is generated by a single \ac{l2r} decoder. This distribution is dependent on the past and thus, has no information about the future context. For this reason,  we extend the model by adding a second \ac{r2l} decoder, which is trained on time-reversed output labels to generate $p(\overleftarrow{y}_l|p_{L:l+1}, \boldsymbol{X})$. The reverse distribution contains beneficial information for the \ac{l2r} decoder since the decoder has no access to future labels. The \ac{r2l} decoder contains an individual attention network, which includes a likewise scoring mechanism as the \ac{l2r} decoder. The decoders learn to create the posterior $p(\overrightarrow{\boldsymbol{y}}|\boldsymbol{X}, \overrightarrow{\theta})$ for the \ac{l2r} and $p(\overleftarrow{\boldsymbol{y}}|\boldsymbol{X}, \overleftarrow{\theta})$ for the \ac{r2l} case, respectively. Thus, $\overrightarrow{\theta}$ represents the attention and decoder parameters for target labels, which are typically time encoded (e.g., \textit{cat}) and $\overleftarrow{\theta}$ are the attention and decoder parameter of the time-reversed target labels (e.g., \textit{tac}).

In an ideal case, the posteriors of both decoders should satisfy the following condition:
\begin{equation}
p(\overrightarrow{\boldsymbol{y}}|\boldsymbol{X}, \overrightarrow{\theta}) = p(\overleftarrow{\boldsymbol{y}}|\boldsymbol{X}, \overleftarrow{\theta}),
\label{eq:posteriors}
\end{equation}
as both networks receive the same amount of information. However, the decoders depend on a different context, i.e., the \ac{l2r} on past context and the \ac{r2l} on future context, which results in a similar but not equal training criterion. 

\subsection{Regularization for Equal Sequence Lengths}

If we apply chars as target values for training the attention model,  we are dealing with equal output sequence lengths since there is no difference between the forward and reverse encoding of a word. Therefore, we extend the loss $\mathcal{L}_{\text{CE}}$ similar to\,\cite{zheng2019forward} with a regularization term to retrieve the global loss $\mathcal{\tilde{L}}$:
\begin{equation}
\mathcal{\tilde{L}} = \alpha \mathcal{L}_{\text{CE}}(\overrightarrow{\theta}) + (1 - \alpha)\mathcal{L}_{\text{CE}}(\overleftarrow{\theta}) + \lambda \Omega(\overrightarrow{\theta}, \overleftarrow{\theta}),
\label{eq:char_regularization}
\end{equation}
where $\alpha$ defines a weighting factor for the losses, and $\Omega(\overrightarrow{\theta}, \overleftarrow{\theta})$ is a regularizer term weighted by $\lambda$. We apply the $L_2$ distance between the decoder outputs $\overrightarrow{\boldsymbol{y}}\in\mathbb{R}^{K}$ and $\overleftarrow{\boldsymbol{y}}\in\mathbb{R}^{L}$ with $K=L$ as regularization. Thus, $\Omega(\overrightarrow{\theta}, \overleftarrow{\theta})$ is defined it as:
\begin{equation}
\Omega(\overrightarrow{\theta}, \overleftarrow{\theta}) = \frac{1}{K} \sum_{k=1}^{K} ||\overrightarrow{y}_k - \overleftarrow{y}_k||_2.
\label{eq:omega}
\end{equation}
The regularization term forces the network to minimize the distance between outputs of the \ac{l2r} and \ac{r2l} decoders. Therefore, the  \ac{l2r} network gets access to outputs that are based on future context information to utilize its knowledge and increase the overall performance.  Note that this kind of regularization is only feasible as we are dealing with equal sequence lengths, which makes it simple to create $\Omega(\overrightarrow{\theta}, \overleftarrow{\theta})$.

\subsection{Regularization for Unequal Sequence Lengths}

We can extend the approach above by applying \ac{bpe} units instead of chars. However, in contrast to chars, we face the problem of obtaining unequal sequence lengths $\overrightarrow{\boldsymbol{y}}\in\mathbb{R}^{K}$ for \ac{l2r} and $\overleftarrow{\boldsymbol{y}}\in\mathbb{R}^{L}$ for \ac{r2l} decoders with $K \neq L$. In fact, we create the same number of \ac{bpe} units, however, they differ between the \ac{l2r} and \ac{r2l} decoders, which results in a difference encoding (e.g., c a t\_ for \ac{l2r} and ta c\_ for the \ac{r2l}). Thus, the proposed regularization in \autoref{eq:omega} is not feasible. We resolve this issue utilizing a differentiable version of the \ac{dtw} algorithm\cite{cuturi2017soft} as a distance measurement between temporal sequences of arbitrary lengths the so-called soft-\ac{dtw} algorithm. By defining a soft version of the \textit{min} operator with a softening parameter $\gamma$:
\begin{equation}
\min^{}{\hspace{-2px}}^\gamma \{ a_1,\cdots,a_n \} \coloneqq
\begin{cases}
\min_{i \leq n}a_i & \gamma = 0\\
-\gamma\log\sum\limits_{i=1}^{n}\text{e}^{a_i / \gamma} & \gamma > 0,	
\end{cases}
\end{equation}
we can rewrite the soft-\ac{dtw} loss as a regularization term $\Omega(\overrightarrow{\theta}, \overleftarrow{\theta})$ similar as above:
\begin{equation}
\Omega(\overrightarrow{\theta}, \overleftarrow{\theta}) = \min^{}{\hspace{-2px}}^\gamma \{ \langle \boldsymbol{A}, \Delta(\overrightarrow{\boldsymbol{y}}, \overleftarrow{\boldsymbol{y}})\rangle, \boldsymbol{A} \in \mathcal{A}_{k,l} \}.
\label{eq:bpe_regularization}
\end{equation}

Here, $\langle\cdot{,}\cdot\rangle$ is the inner product of two matrices, $\boldsymbol{A}$ is an alignment matrix of a set $\mathcal{A}_{k,l} \subset \{0,1\}^{k,l}$ which are binary matrices that contain paths from $(1,1)$ to $(k,l)$ by only applying $\downarrow$, $\rightarrow$ and $\searrow$ moves through this matrix and ${\Delta(\overrightarrow{\boldsymbol{y}}, \overleftarrow{\boldsymbol{y}}) \coloneqq [\delta(\overrightarrow{y}_k,\overleftarrow{y}_l)]}$ is defined by a distance function $\delta(\overrightarrow{y}_k,\overleftarrow{y}_l)$ (e.g., Euclidean distance). Based on the inner product, we retrieve an alignment cost for all possible alignments between $\overrightarrow{\boldsymbol{y}}$ and $\overleftarrow{\boldsymbol{y}}$. Since we force the network to also minimize $\Omega(\overrightarrow{\theta}, \overleftarrow{\theta})$, it has to learn a good match between the different sequence lengths of the \ac{l2r} and \ac{r2l} decoders.

\section{Experiments}

\begin{table*}[htbp]
	\centering
	\caption{Evaluation of our approach on TEDLIUMv2 and LibriSpeech with the resulting \acp{wer} (\%) for all five setups}
	\begin{tabular}{p{5.91em}cccccccccccc}
		\toprule
		\multicolumn{1}{r}{} & \multicolumn{4}{c}{TEDLIUMv2\cite{rousseau2014enhancing}} & \multicolumn{8}{c}{LibriSpeech\cite{panayotov2015librispeech}} \\
		\cmidrule(lr){2-5} \cmidrule(lr){6-13}    \multicolumn{1}{r}{} & \multicolumn{2}{c}{char} & \multicolumn{2}{c}{BPE} & \multicolumn{4}{c}{char} & \multicolumn{4}{c}{BPE} \\
		\cmidrule(lr){2-3} \cmidrule(lr){4-5} \cmidrule(lr){6-9} \cmidrule(lr){10-13} \multirow{2}{*}{\hfil Methods} & \multirow{2}{*}{\hfil dev} & \multirow{2}{*}{\hfil test} & \multirow{2}{*}{\hfil dev} & \multirow{2}{*}{\hfil test} & \multicolumn{1}{P{2.365em}}{dev-clean} & \multicolumn{1}{P{2.365em}}{\hfil dev-\newline{}other} & \multicolumn{1}{P{2.365em}}{\hfil test-\newline{}clean} & \multicolumn{1}{P{2.365em}}{\hfil test-\newline{}other} & \multicolumn{1}{P{2.365em}}{\hfil dev-\newline{}clean} & \multicolumn{1}{P{2.365em}}{\hfil dev-\newline{}other} & \multicolumn{1}{P{2.365em}}{\hfil test-\newline{}clean} & \multicolumn{1}{P{2.365em}}{\hfil test-\newline{}other} \\
		\midrule
		\multicolumn{1}{l}{Forward} & 16.77 & 17.32 & 17.83 & 18.00 & 7.69 & 20.67 & 7.72 & 21.63 & 7.59 & 20.98 & 7.67 & 21.92 \\
		\multicolumn{1}{l}{Backward} & 18.12 & 18.47 & 18.57 & 17.99 & 7.60 & 20.78 & 7.54 & 21.83 & 7.53 & 20.94 & 7.60 & 21.71 \\
		\multicolumn{1}{l}{Backward Fixed} & 23.34 & 23.77 & 25.55 & 25.01 & 11.39 & 28.36 & 11.75 & 28.53 & 12.07 & 28.63 & 12.39 & 29.06 \\
		\multicolumn{1}{l}{Dual Decoder} & 16.47 & 17.12 & 17.70 & 18.08 & 7.29 & 20.99 & 7.60 & 22.00 & 7.46 & 21.29 & 7.70 & 22.01 \\
		\multicolumn{1}{l}{Dual Decoder Reg} & \textbf{15.68} & \textbf{15.94} & \textbf{16.75} & \textbf{17.42} & \textbf{7.24} & \textbf{19.96} & \textbf{7.02} & \textbf{20.95} & \textbf{7.17} &  \textbf{20.01} &  \textbf{7.33} &  \textbf{20.63} \\
		\bottomrule
	\end{tabular}%
	\label{tab:results}%
	\vspace{-0.3cm}
\end{table*}%

\subsection{Training Details}
All our experiments are evaluated on the smaller dataset TEDLIUMv2\,\cite{rousseau2014enhancing} and the larger dataset LibriSpeech\,\cite{panayotov2015librispeech}. TEDLIUMv2 has approximate 200\,h of training data, whereas LibriSpeech contains of 960\,h of training data.

We preprocess both datasets by extracting 80-dimensional log Mel features and adding the corresponding pitch features, which results in an 83-dimensional feature vector. Furthermore, we apply chars and \ac{bpe} units as target labels. The chars are directly extracted from the datasets, whereas the \ac{bpe} units are created by a language-independent sub-word tokenizer. For all experiments, we select 100 \ac{bpe} units, which seem sufficient for our approach. 

The proposed architecture is created in the ESPnet toolkit\,\cite{watanabe2018espnet} and trained by the standard training script for attention models. The encoder is built up by four \ac{blstmp} layers of dimension 1024. Each decoder contains a single \ac{lstm} network with 1024 cells and a linear output layer.

We perform a three-stage training scheme inspired by \cite{zheng2019forward}. In the first stage, we train a standard attentional network with a \ac{l2r} decoder. Then, we apply the pretrained encoder, freeze its weights and train the \ac{r2l} model. Finally, we combine both networks into one model to receive the final architecture. In all stages, we optimize the network with Adadelta initialized with an $\epsilon = 10^{-8}$. If we do not observe any improvement of the accuracy on the validation set, we decay $\epsilon$ by a factor of $0.01$ and increment a patience counter by one. We apply an early stopping of the training if the patient counter exceeds three. The batch-size is set to 30 for all training steps.

Depending on the target labels in the third training stage, i.e., chars or \ac{bpe} units, we deploy two different techniques to regularize the \ac{l2r} decoder. For chars, forward sequences $\overrightarrow{\boldsymbol{y}}$ and backward sequences $\overleftarrow{\boldsymbol{y}}$ have equal lengths. Thus, we add a $L_2$ regularizer identical to \autoref{eq:char_regularization} and scale it with $\lambda = 1$ for the smaller and $\lambda = 0.1$ for the bigger dataset. On the other hand, for \ac{bpe} units, we utilize the soft-DTW from \autoref{eq:bpe_regularization} as a regularizer since it represents a distance measurement between the unequal sequence lengths $\overrightarrow{\boldsymbol{y}}$ and $\overleftarrow{\boldsymbol{y}}$, which we want to minimize. Here, we set $\gamma = 1$ and scale the regularization with $\lambda = 10^{-4}$ for both datasets.
Besides the added regularizations for chars and \ac{bpe} units, we regularize the \ac{l2r} network further by applying $\alpha = 0.9$ in all the experiments. Thereby, we ensure that the overall training is focused on the \ac{l2r} decoder network. 
During decoding, we remove the \ac{r2l} network since it is only necessary in the training stages. As a result, we are not changing or adding complexity to the final model during decoding.

\subsection{Benchmark Details}

We evaluate our approach on five different setups. In the \textit{Forward} setup, a model is trained with a standard \ac{l2r} decoder, which is the baseline for all experiments. The second setup is the \textit{Forward} setup, where a model is trained on time-reversed target labels, which results in a \ac{r2l} decoder. We perform a similar approach in \textit{Backward Fixed}, however, we apply the pretrained encoder from the \ac{l2r} model and freeze its weights during training. To solely investigate the effect of the \ac{r2l} decoder as regularization, we define the \textit{Dual Decoder} setup. The model consists of a shared encoder from \textit{Forward} and the pretrained \ac{l2r} and \ac{r2l} decoder from the \textit{Forward} and \textit{Backward Fixed} setups. The combined model is trained with $\alpha = 0.9$ and $\lambda=0.0$. In the last \textit{Dual Decoder Reg} setup, which is similar to the \textit{Dual Decoder} setup, we include the $L_2$ distance\,\cite{zheng2019forward} for chars and the soft-\ac{dtw} loss\,\cite{cuturi2017soft} for \ac{bpe} units as target labels.
Instead of performing the forward and backward beam search as in\,\cite{mimura2018forward}, we only apply a forward beam search deploying the \ac{l2r} decoder with a beam size of 20.

\subsection{Results}

In \autoref{tab:results}, we present the results of our approach applying chars and \ac{bpe} units for the TEDLIUMv2\cite{rousseau2014enhancing} and LibriSpeech\,\cite{panayotov2015librispeech} datasets. 

For the smaller dataset TEDLIUMv2, we observe a clear difference in \acp{wer} between the \textit{Forward} and the \textit{Backward} setup. Ideally, the performance of these setups should be equal, as both networks receive the same amount of information. However, we observe an absolute difference of 1\% \ac{wer} for all evaluation sets, except for the test \ac{bpe} set. One explanation for this variation may be that the \textit{Backward} setup is more complex. Since the dataset contains only around 220\,h of training data, the number of reverse training samples could not be sufficient. In the bigger dataset LibriSpeech, the first two setups obtain nearly the same \ac{wer} with only a minor difference. This dataset contains nearly five times the data of the smaller dataset and therefore, the network in the \textit{Backward} setup receives enough reverse training examples. It seems, that the amount of data seems crucial for the \ac{r2l} decoder to satisfy \autoref{eq:posteriors}.

In the \textit{Backward Fixed} setup, we can verify the strong dependency of the decoder, relying on the high-level representation of features created by the encoder. Although we do not change the information of the target labels by reversing them, the fixed encoder from the \textit{Forward} setup learned distinct, high-level features, which are based on past context. We observe this by a decline of the \acp{wer} in both datasets. Even though, the utilized \acp{blstmp} in the encoder network receive the complete feature sequence in the input space, they generate high-level features based on past label context, since they do not have access to future labels. As a result, the \ac{r2l} model applying a fixed encoder from the \textit{Forward} setup is worse compared to the trainable encoder in the \ac{r2l} model. 

In the \textit{Dual Decoder} setup, we follow the idea of \cite{mimura2018forward} to apply the \ac{r2l} model as a regularizer of the \ac{l2r} network. Interestingly, the \ac{r2l} decoder is not able to effectively support the \ac{l2r} decoder. We recognize only a slight improvement of the \ac{wer}, which is not consistent in both datasets. Therefore, a simple weighting of the loss during training is not sufficient to enhance the \ac{l2r} decoder. One reason might be that the \ac{l2r} decoder receives only implicit information from the \ac{r2l} decoder by weighting the losses, which is considered not valuable for the optimization of the \ac{l2r} decoder.

To induce valuable information, we add our proposed regularization terms in the last \textit{Dual Decoder Reg} setup. The overall network is forced to minimize the added regularization terms explicitly. The \ac{l2r} decoder can directly utilize information of the \ac{r2l} decoder to improve its predictions. We receive the overall best \ac{wer} for the last setup. For the TEDLIUMv2 dataset, we recognize an average relative improvement of 7.2\% for the char and 4.4\% for the \ac{bpe} units. For the LibriSpeech dataset, we are able to receive an average relative improvement of 4.9\% for the char and 5.1\% for the \ac{bpe} units.

Compared to other state-of-the-art approaches, we decided not to include \ac{ctc} and a language model since their integration into our approach raises several issues as we deal with a shared encoder and unequal sequence lengths among the two decoders.


\section{Conclusion}

Our work presents a novel way to integrate a second decoder for attention models during the training phase. The proposed regularization terms support the standard \ac{l2r} model to utilize future context information from the \ac{r2l} decoder, which is usually not available during optimization.
We solved the issue of regularizing unequal sequence lengths, which arise applying \ac{bpe} units as target values, by adding a soft version of the \ac{dtw} algorithm. We outperform conventional attention models independent of the dataset size. Our regularization technique is simple to integrate into a conventional training scheme, does not change the overall complexity of the standard model, and only adds optimization time. 

\bibliographystyle{IEEEtran1}
\bibliography{mybib}

\end{document}